\documentclass[12pt,preprint]{aastex}
\newcommand{\nraoblurb}{The National Radio Astronomy Observatory is
a facility of the National Science Foundation operated under cooperative
agreement by Associated Universities, Inc.}
\newcommand{\HII}{{\rm H\,}{{\sc ii}}}       
\newcommand{\GHII}{{\rm GH\,}{{\sc ii}}}       
\newcommand{\HI}{{\rm H\,}{{\sc i}}}       
\newcommand{\etal}{et al.}       
\newcommand{\cmthree}{cm$^{-3}$}       
\newcommand\kms{km~s$^{-1}$}
\catcode`\@=11
\newcommand{\gsim}{${\mathrel{\mathpalette\@versim>}}$}
\newcommand{\lsim}{${\mathrel{\mathpalette\@versim<}}$}
\newcommand{\@versim}[2]{\lower 2.9truept \vbox{\baselineskip 0pt \lineskip
    0.5truept \ialign{$\m@th#1\hfil##\hfil$\crcr#2\crcr\sim\crcr}}}
\catcode`\@=12
\newcommand{\Msun}{$\ensuremath\,M_\odot$}

\slugcomment{Accepted by ApJ}

\shorttitle{H and He RRLs toward Luminous WMAP source}
\shortauthors{D. Anish Roshi et al. }

\begin{document}

\title{On the Ionization of Luminous WMAP Sources in the Galaxy :
Constraints from He Recombination Line Observations with the GBT}

\author{D. Anish Roshi}
\affil{National Radio Astronomy Observatory (NRAO)\footnote{\nraoblurb}, Green Bank, WV 24944, 
and NRAO Technology Center, Charlottesville, VA 22903-4608, aroshi@nrao.edu}

\author{Adele Plunkett}
\affil{Department of Astronomy, Yale University, P.O. Box 208101, New Haven CT 06520, adele.plunkett@yale.edu}

\author{Viviana Rosero}
\affil{
New Mexico Institute of Mining and Technology, Physics Department, 801 Leroy Place, Socorro, NM 
87801, viviana@nmt.edu}

\and

\author{Sravani Vaddi}
\affil{Astrophysical Sciences and Technology, Rochester Institute of Technology, NY 14623, sxv1249@rit.edu}

\begin{abstract}
The Wilkinson Microwave Anisotropy Probe (WMAP) free-free 
foreground emission map is used to identify diffuse ionized 
regions (DIR) in the Galaxy (Rahman \& Murray 2010). It has been 
found that the 18 most luminous WMAP sources produce more than
half of the total ionizing luminosity of the Galaxy. We observed radio
recombination lines (RRLs) toward the luminous WMAP source G49.75$-$0.45 with the Green Bank Telescope
near 1.4 GHz. Hydrogen RRL is detected toward the source but no
helium line is detected, implying that $n_{He^+}/n_{H^+} < 0.024$. This limit
puts severe constraint on the ionizing spectrum. The total ionizing luminosity of G49 
($3.05 \times 10^{51}$ s$^{-1}$) is $\sim$ 2.8 times the luminosity
of all radio \HII\ regions within this DIR and this is generally the
case for other WMAP sources. Murray \& Rahman (2010) 
propose that the additional ionization is due to massive clusters 
($\sim 7.5 \times 10^3$ \Msun\ for G49) embedded in the WMAP sources. Such
clusters should produce enough photons with energy $\ge$ 24.6 eV to fully
ionize helium in the DIR. Our observations rule out a simple model with G49 ionized by a massive cluster. We also considered `leaky' \HII\ region models for the ionization of the DIR, suggested by  Lockman and Anantharamaiah, but these models also cannot explain our observations. 
We estimate that the helium ionizing photons need to be attenuated by \gsim\ 10 times 
to explain the observations. 
If selective absorption of He-ionizing photons by dust is causing this additional
attenuation, then the ratio of dust absorption cross sections for He- and H-ionizing 
photons should be \gsim 6.
\end{abstract}


\section{Introduction}

The existence of a diffuse ionized gas in the Galaxy is evident
from a variety of observations (\nocite{he63}Hoyle \& Ellis 1963,
see review by \nocite{hetal09}Haffner et al. 2009). This gas, 
referred to as the Warm Ionized
Medium (WIM), is now considered as one of the major components of 
the interstellar medium. The WIM has been primarily studied using
optical emission lines, which indicates that it contains 90\% of 
the ionized mass of the Galaxy (\nocite{hetal09}Haffner \etal \ 2009). These observations
also indicate that the local electron density of the WIM is in the range 0.01 
to 0.1 \cmthree. In the inner galaxy, the optical lines suffer
extinction and hence the distribution of the
ionized gas has been studied in the radio frequency regime.
In particular, low frequency (\lsim\ 2 GHz) RRL observations have been used to detect diffuse ionized regions (DIRs) with local density in the
range 1 to 10 \cmthree. In the literature, this DIR is referred 
to as Galactic Ridge RRL emission by \nocite{dmp72}
Davies, Matthews \& Pedlar (1972) (see also \nocite{gg70}Gottesman \& 
Gordon 1970), extended low-density
ionized gas by \nocite{m78}Mezger (1978), evolved \HII\ region
by \nocite{s76}Shaver (1976),  \HII\ envelopes (i.e.
low-density envelopes of compact \HII\ regions) by 
\nocite{l76}Lockman (1976) and \nocite{a86}Anantharamiah (1986)
(see also \nocite{ra00}Roshi \& Anantharamaiah 2000)
and more recently as extended low-density warm ionized medium 
by \nocite{pb93}Petuchowski \& Bennett (1993) and \nocite{h94}Heiles (1994). 
The latter authors consider the DIR as a higher density version
of the WIM in the inner Galaxy, though the exact relationship between
the DIR and WIM is still uncertain. 

In addition to spectral lines, DIRs have been studied through radio free-free emission  (e.g. \nocite{m78}Mezger 1978). 
More recently, \nocite{mr10}Murray \& Rahman (2010) used the 
free-free galactic foreground emission model, obtained from 
the Wilkinson Microwave Anisotropy Probe (WMAP) data, 
to study the DIR. For the WMAP measurements of Cosmic Microwave Background (CMB) 
emission, the galactic foreground is accurately characterized using
a maximum entropy model (MEM; \nocite{betal03}Bennett \etal\ 2003, \nocite{getal09}
Gold \etal\ 2009). The foreground is modeled as a linear 
combination of emission due to synchrotron, free-free and vibrational
excitation of dust with power law frequency dependence. The MEM
finally produces sky models for each of these emission processes.
Since the free-free model has an angular resolution of $\sim$ 1 $\deg$ (after
smoothing), it is more sensitive to emission from DIRs compared to compact \HII\ regions.
\nocite{mr10}Murray \& Rahman (2010) used this free-free model to identify
emission regions with 94 GHz flux density \gsim 10 Jy. By considering that the
WMAP sources are associated with known \HII\ regions, which are either
cataloged by \nocite{r03}Russeil (2003) or identified in the
Midcourse Space Experiment (MSX; \nocite{petal01}Price \etal\ 2001) or 
Spitzer GLIMPSE (\nocite{cetal06}Churchwell \etal\ 2006) data set, distances can be assigned. The free-free flux density, source angular size and 
the distance are used to estimate the Lyman continuum (Lyc) luminosity 
required for ionization balance. They found that the 18 most luminous
WMAP sources contribute half of the total ionizing luminosity 
($3.2 \times 10^{53}$ s$^{-1}$) of the Galaxy.

\nocite{rm10}Rahman \& Murray (2010) studied in detail a subset of the luminous 
WMAP sources. The estimated Lyc luminosities of these sources
are in the range $10^{51.5}$ to $10^{52}$ s$^{-1}$. They
show that the radio \HII\ regions embedded in these WMAP sources
contribute only $\frac{1}{2}$ to $\frac{1}{3}^{rd}$ of the Lyc luminosity obtained from 
the WMAP emission. \nocite{mr10}Murray \& Rahman (2010) suggest that the
additional ionization is due to embedded stellar clusters. The
mass of the cluster required to produce the ionizing luminosity is
\gsim\ $10^4$ \Msun. These clusters evaded detection primarily due to
heavy dust obscuration in the star forming region. 

For the most widely used initial mass function (IMF e.g.
\nocite{s86}Scalo 1986 or \nocite{metal02}Muench \etal\ 2002)
of the stellar content of clusters, the bulk of the ionizing 
photons occurs for star with mass $\sim$ 40 \Msun
(equivalent to O5 or earlier). These stars should produce enough
photons with energy $\ge$ 24.6 eV so that most of the helium (He) in the DIR
should have been ionized. In fact, \nocite{hetal96}Heiles \etal\ (1996)
searched for He RRLs from the DIR and did not detect them. 
\nocite{hetal96}Heiles \etal\ (1996) found that the $spatial$ average is
$n_{He^+}/n_{H^+} < 0.013$. This limit is about 7.7 times lower than the expected value 
of 0.1 from cosmic abundance of He, if all the He is in He$^+$. We note here that
the He\,{\sc i} $\lambda$5876 observations made toward the WIM also show 
$n_{He^+}/n_{H^+}$ \lsim 0.027 (\nocite{rt95}Reynolds \& Tufte 1995; 
we have taken $n_{H^+}/n_{H} \sim$ 1 for the WIM).  
  
In this paper, we present results of 1.4 GHz RRL observations toward G49.75$-$0.45, located within the luminous WMAP source G49 (\nocite{rm10}Rahman \& Murray 2010), 
with the National Radio Astronomy Observatory
(NRAO) Green Bank Telescope (GBT). 
Our objective was to do a deep integration toward a single pointing in G49 (i.e. without spatial average) away from bright radio \HII\ regions and determine $n_{He^+}/n_{H^+}$. 
The observations and data analysis procedure
are presented in Section~\ref{obs}. As discussed in Section~\ref{result},
we have detected hydrogen (and carbon) 
RRL toward G49.75$-$0.45 but no He line is detected. The upper limit 
implies that $n_{He^+}/n_{H^+} < 0.024$. The implication of this upper limit on the 
spectrum of the ionizing source is discussed in Section~\ref{disc}.
The main results of the paper is summarized in Section~\ref{sum}.

\section{Observing Strategy, RRL Observations and Data reduction}
\label{obs}

We selected the position G49.75$-$0.45, located toward the WMAP source G49 (\nocite{rm10}Rahman \& Murray 2010), for observations.
The WMAP source is located in galactic longitude away from other 
dense star forming regions in the inner Galaxy and is near the 
outer edge of Sagittarius-Carina arm.
The star forming complex W51 is located within this source. 
We selected the position in G49 (see Fig.~\ref{fig1}a) such that it is away from compact \HII\ regions and hence samples the diffuse thermal gas seen by WMAP.

The RRL observations are made with the GBT. The frequency band for
the observations is chosen as follows.
The approximate average continuum brightness temperature toward G49
at 4.785 GHz (\nocite{aetal79}Altenhoff et al. 1979) and 408 MHz 
(\nocite{hetal82}Haslam et al. 1982) are 
respectively $\sim$ 0.5 K and $\sim$ 150 K. Assuming that the emission at 
408 MHz is dominated by non-thermal emission and has a spectral index 
of $-$2.6, the expected non-thermal
contribution at 4.785 GHz is $\sim$ 0.25 K. An emission
measure of $\sim$ 1800 pc cm$^{-6}$ for the ionized gas is estimated to produce the
thermal emission of 0.25 K at 4.785 GHz. We take the electron temperature of 7000 K, 
same as that of the compact \HII\ region, for this calculation 
(\nocite{detal80}Dowens \etal\ 1980). For the
estimated emission measure and a mean electron density of $\sim$ 5 \cmthree\ 
(\nocite{ra01}Roshi \& Anantharamaiah 2001), the expected
signal to noise ratio of RRL detection peaks near 1 GHz. Taking into account 
RFI near 1 GHz, we decided to observe with the L-band (1.1 to 1.75 GHz) system of the GBT.  
 
Observations were made with the GBT on 13 July 2011.
The L-band receiver is a single-beam, dual-linear-polarization
system. The FWHM beam width at this frequency is $\sim$9\arcmin.
The GBT spectrometer was used to simultaneously observe 
the 169$\alpha$, 168$\alpha$, 167$\alpha$ and 166$\alpha$ transitions 
of hydrogen (H), helium and carbon from both polarizations. A 
bandwidth of 12.5 MHz and 8192 channels for the spectrometer were selected, 
which gave a spectral resolution of $\sim$1.5 KHz ($\sim\ 0.3$\kms).
This higher spectral resolution was helpful in editing out narrow band RFI.
We employed a dual-Dicke frequency switching to measure the reference
spectrum. The frequency is switched by 2.5 MHz, which corresponds to
$\sim$526 \kms. All the spectral lines of interest fall within this velocity range.
The front-end bandwidth was restricted between 1.3 and 1.45 GHz
to reduce the effect of out of band RFI. The total on-source observing time
was about 3.5 hrs. The calibrator 3c295 was observed in the
beginning for pointing, focus and flux calibration.

The data analysis is done in GBTIDL and Matlab/Octave. Data 
corresponding to all the transitions and the two polarizations are 
examined for RFI and averaged separately in GBTIDL. 
The \HI\ 21cm line, present in the 166$\alpha$ band, was 
examined and found not to affect the RRL.  
Online Doppler correction could be done only for the 166$\alpha$ transition
and so the residual corrections for other transitions were done
by FFT shift method in GBTIDL before averaging. A 3rd order baseline
was fitted to these spectra using data points from the spectral 
region free of RRLs. The velocity
resolution of the spectra of different transitions were then
made identical by FFT re-sampling (see \nocite{retal05}Roshi \etal\ 2005). The 
averaged spectra corresponding to each transition were further edited 
for bad spectral baseline and RFI. The spectrum corresponding to 
the 168$\alpha$ transition had narrow band RFI near the 
expected position of the He line and hence was edited out. 
The spectral baseline of the X-polarization of the 166$\alpha$ transition 
had a ripple, which was also excluded from further analysis. 
Narrow band RFI, which were located at frequencies far
from the expected line frequencies, were edited out using a channel
weighting scheme (\nocite{ra00}Roshi \& Anantharamaiah 2000). 
The edited spectra were averaged to obtain the final integrated spectrum. 
The effective integration time of the final spectrum is about 16.4 hrs
and average $T_{sys}$ is 22.6 K. The spectral amplitudes are calibrated
using the noise cal and hence are given in antenna temperature.
To convert to main beam brightness temperature these values need
to be multiplied by 1.14, since the observed source is extended and
roughly fills the beam. 

\section{Result}
\label{result}

\begin{figure}
\begin{tabular}{cc}
\includegraphics[width=3.0in, height=2.5in, angle=0]{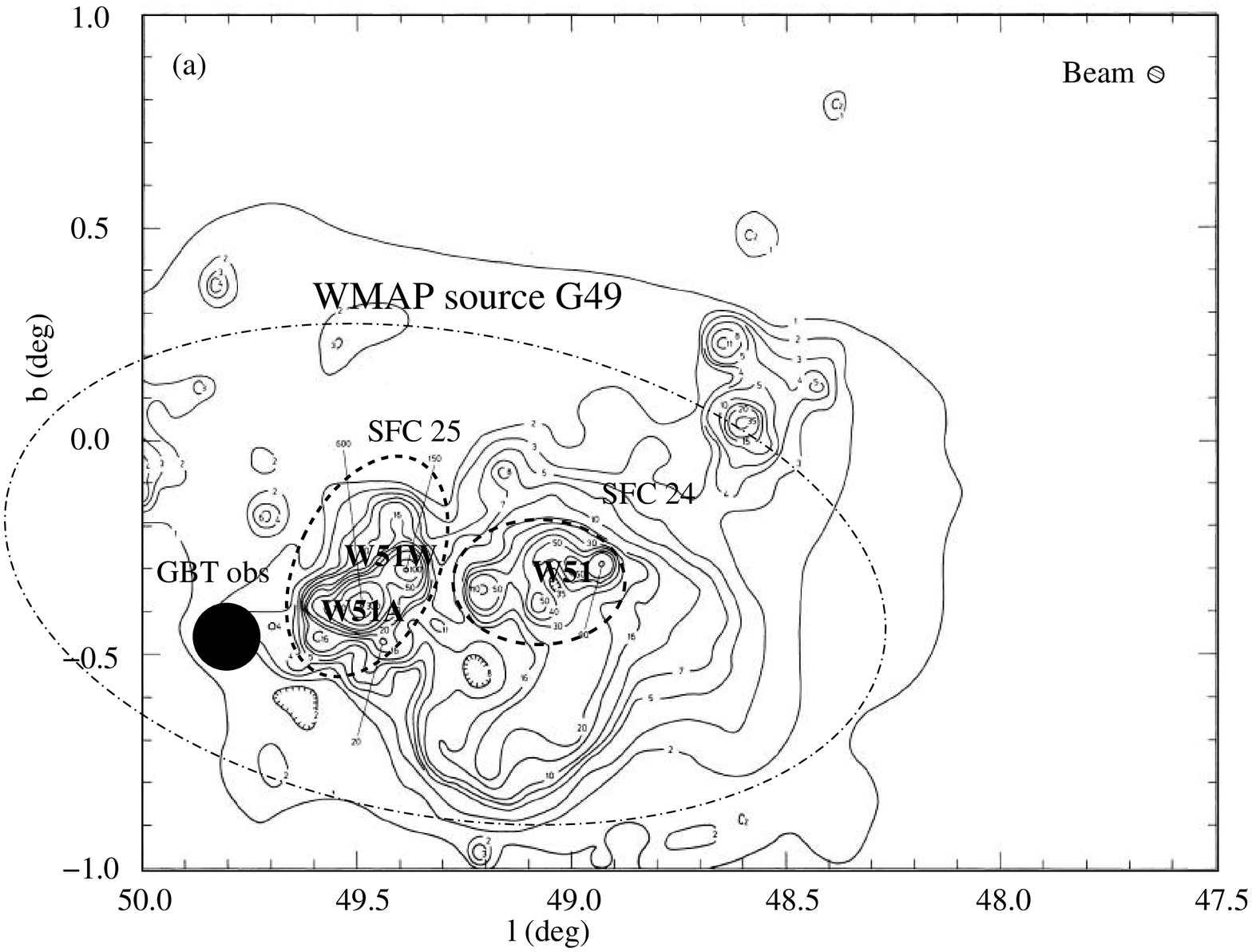} &
\includegraphics[width=3.0in, height=2.5in, angle=0]{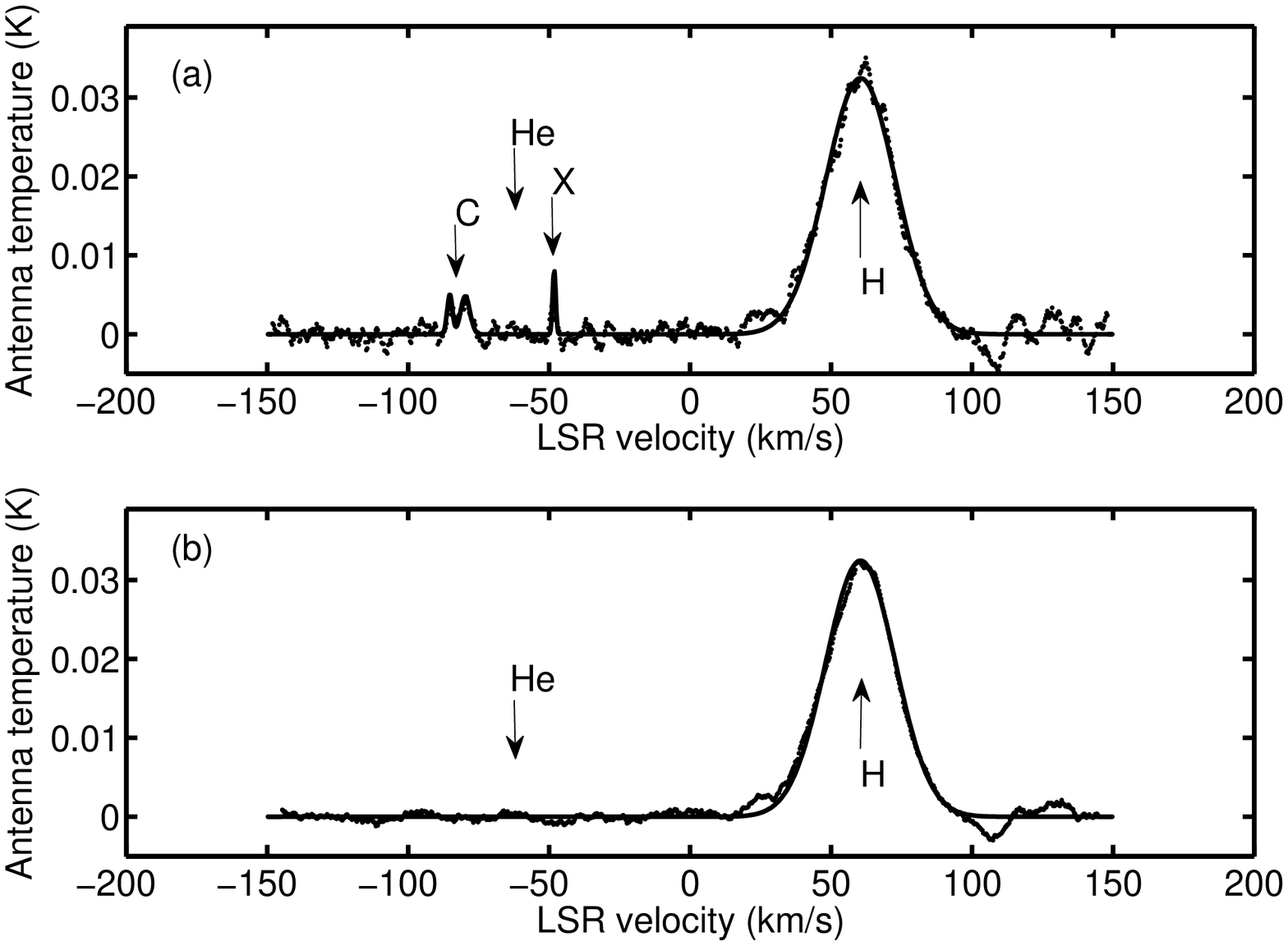} \\
\end{tabular}
\caption{(a) Position (G49.75$-$0.45) observed with the GBT is shown by the 
filled circle on the 4.785 GHz continuum map (Altenhoff et al. 1979). The size of 
the filled circle indicates the GBT beam ($\sim$ 9\arcmin) near 1.4 GHz. 
The WMAP source G49 is shown by the dash-dot ellipse. The giant \HII\ regions 
W51, W51A and W51W are marked. Rahman \& Murray's (2010) starforming complexes 
(SFC, which are super-sets of \HII\ regions in the region) 25 and 24 are also shown. 
(b) The average spectrum obtained toward G49.75$-$0.45. 
The hydrogen and carbon RRLs (and a line feature corresponding to an unknown 
transition marked as X) are clearly detected. The helium line is not detected; 
the upper limit implies that $n_{He^+}/n_{H^+} < 0.024$. The upper limit to 
helium line is determined from the spectrum shown in (c), which is obtained after 
subtracting Gaussian models for the carbon and X lines
from the average spectrum and smoothing to $\sim$ 10.5 \kms\ velocity resolution. 
The velocity range near 110 \kms\ is affected by RFI. \label{fig1}}
\end{figure}

The final spectrum obtained from the data is shown in Fig~\ref{fig1}b. 
The H and carbon RRLs are clearly detected. 
A line feature corresponding to an unknown transition,
marked as X, is also detected. The parameters for Gaussian models of the detected lines 
are given in Table~\ref{tab1}. He line is not detected. To 
determine the upper limit to the He line intensity, Gaussian
models for the carbon and X lines are removed from the integrated spectrum and
the resultant spectrum is smoothed to $\sim$ 10.5 \kms\ velocity resolution. 
The smoothed spectrum is shown in Fig~\ref{fig1}c. The root mean square (RMS)  
of the spectral values over the velocity range $-$120 to 0 \kms\ 
is 4.4 $\times$ 10$^{-4}$. Since
there are only 11 points for estimating the RMS, we have taken the upper limit
for the He line as the 99.99\% confidence limit of the spectral mean
(obtained using Student's t-distribution), which is listed in Table~\ref{tab1}.  
Using this upper limit we determine that $n_{He^+}/n_{H^+} < 0.024$.  
We have neglected the effect of the smaller expected thermal line width
for He compared to H to obtain this ratio and have also considered that H is
fully ionized in the DIR (\nocite{ra01}Roshi \& Anantharamaiah 2001). 

\section{Discussion}
\label{disc}

Three `Giant' \HII\ ( \GHII ) regions are located within the WMAP source G49. 
They are W51, W51A and W51West, all of which are well studied in radio 
and infrared wavelengths (e.g. \nocite{cc04}Conti \& Crowther 2004). The \GHII\ regions
are observed as compact sources in radio continuum images (see Fig~\ref{fig1}a). 
They are located at a kinematic distances of 5.5 kpc (\nocite{r03}Russeil 2003). 
RRL observations near 327 MHz have detected lines toward W51, which arise 
from the DIR since most of the compact \HII\ regions become optically thick 
at this frequency (e.g. \nocite{ra01}Roshi \& Anantharamaiah 2001). The LSR velocity of the
327 MHz RRLs coincide with those detected from the compact \HII\ regions 
at higher frequencies. Therefore in this paper we consider the distance to 
the WMAP source G49 as 5.5 kpc. 

The total Lyc luminosity of \GHII\ regions after dust extinction 
correction is 1.1 $\times 10^{51}$ s$^{-1}$ (\nocite{cc04}Conti \& Crowther 2004).
On the other hand, the total Lyc luminosity obtained from
WMAP free-free emission after dust extinction correction and using a distance
of 5.5 kpc is 3.05 $\times 10^{51}$ s$^{-1}$ (\nocite{mr10}Murray \& Rahman 2010). 
The larger ionizing photon requirement obtained from DIR tracers is
noted earlier (e.g. \nocite{m78}Mezger 1978, \nocite{mw97}McKee \& Williams 1997). As
suggested by \nocite{l76}Lockman (1976) and \nocite{a86}Anantharamiah (1986), 
most of the Lyc photons from \HII\ regions may leak out to produce 
low-density envelopes which are observed as the DIR 
(see also \nocite{mw97}McKee \& Williams 1997). 
In this picture, the ionizing sources are the same star clusters embedded in the 
\GHII\ regions (\nocite{cc04}Conti \& Crowther 2004). The two ionizing luminosities 
estimated above for G49 imply that $\sim$ 63 \% of the Lyc photons 
are leaking out of the \GHII\ regions. 
\nocite{mr10}Murray \& Rahman (2010) suggested another possibility that the DIR
are ionized by embedded massive clusters not associated with classical radio
\HII\ regions within the WMAP sources. Such clusters are not directly detected 
presumably due to high obscuration. 

\begin{figure}
\includegraphics[height=5in, angle=0]{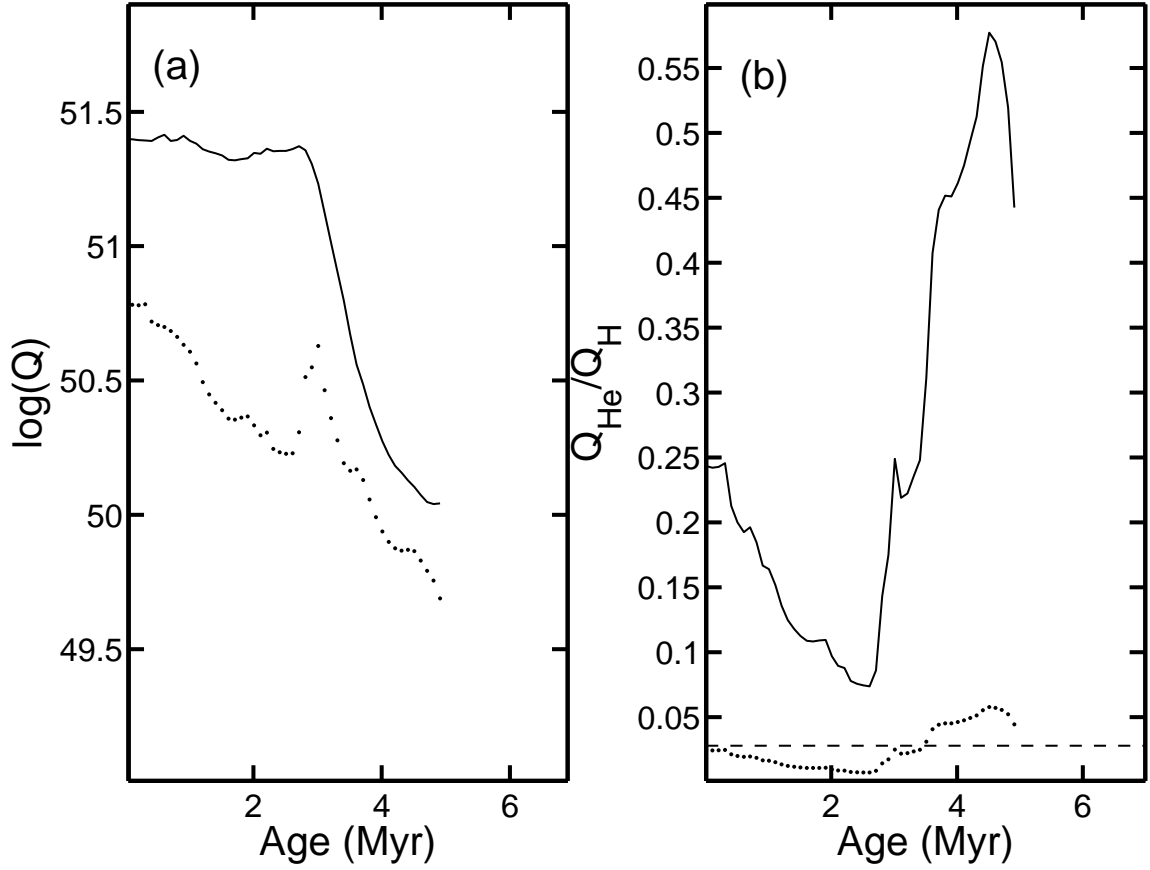}
\caption{(a) Log of $Q_H$ (solid line) and $Q_{He}$ (dotted line) vs 
age of a $7.5 \times 10^3$ \Msun\ star cluster obtained using Starburst99.  
Muench IMF (Muench et al. 2002) and solar metallicity are used for the 
calculation. (b) The ratio $Q_{He}/Q_{H}$ vs age obtained for the 
$7.5 \times 10^3$ \Msun\ star cluster. The upper limit on this ratio determined from
our RRL observation is marked with the dashed line. The dotted line shows the 
ratio $Q_{He}/Q_{H}$ with the photons of energy $\ge$ 24.6 eV attenuated 
by a factor of 10 relative to the current Starburst99 model prediction. \label{fig2}}
\end{figure}

Here we explore the possibility of using He RRL observations
toward the G49 region to distinguish between these different possible
ionization and to constrain the spectrum of the ionizing photons. 
Since the ionization potential of He (24.6 eV) is greater than that of
H (13.6 eV), the observed He to H line ratio
can be used to constrain the ionizing spectrum.  We calculate the
ionizing photon from a stellar cluster using Starburst99 (
\nocite{letal99}Leitherer et al. 1999, \nocite{vl05}V\'{a}zquez \& Leitherer 2005).
The mass of the cluster to get the required total 
Lyc photons for G49 is $\sim 7.5 \times 10^3$ \Msun; the IMF used
for this calculation is from \nocite{metal02}Muench et al. (2002) 
\footnote{The effects of 
stochasticity on the luminosity of cluster with mass a few times $10^3$ \Msun will
be investigated in a future publication.}.
In Murray \& Rahman's picture, a single cluster will have this 
mass. On the other hand in the \nocite{l76} Lockman (1976) and \nocite{a86} Anantharamaiah (1986) pictures 
the mass gets distributed to three clusters 
in the three \GHII\ regions. In either case, the ratio of He ionizing photons, 
$Q_{He}$, (photon energy interval 24.6 to 54.4 eV) to the H ionizing photons, $Q_H$,
(photon energy interval 13.6 to 24.6 eV) is $\sim$ 0.24 (see Fig~\ref{fig2}). 
From the ionization model of \HII\ region this $Q_{He}/Q_H$ 
implies that the He and H Str\"{o}mgren spheres overlap 
(\nocite{m71}Mathis 1971) and hence He line should
have been detected. However, our observations did not detect He line in G49. 

The $n_{He^+}/n_{H^+}$ ratio obtained from high frequency RRL observations toward 
compact \HII\ regions in G49 is $>$ 0.066 (\nocite{cmh74}Churchwell, 
Mezger \& Huchtmeier 1974, \nocite{lrc79}Lichten, Rodriguez, 
Chaisson 1979, \nocite{tmp80}Thum, Mezger, Pankonin 1980, 
\nocite{mn81}McGee \& Newton 1981, \nocite{m94}Mehringer 1994, 
\nocite{betal11}Bell \etal\ 2011). The low observed $n_{He^+}/n_{H^+}$ 
for the DIR in G49 thus indicates that the He ionizing photons are 
eliminated on scales somewhat larger than the compact \HII\ regions. 
If the DIR is ionized by 
clusters embedded in the \GHII\ regions, as suggested by 
\nocite{l76} Lockman (1976) and \nocite{a86} Anantharamaiah (1986), 
then one possibility is that the spectrum of the photons 
changes while diffusing out of the \HII\ regions. However,
models of `leaky \HII\ regions' show that the He ionizing photons
can be significantly suppressed only for low leakage ($\sim$ 15\%)
and for effective temperature of ionizing source $<$ 45000 K
(\nocite{hw03}Hoopes \& Walterbros 2003, \nocite{wm04}Wood \& Mathis 2004); both these
conditions are not true for G49. 

\nocite{mr10}Murray \& Rahman (2010) identifies stellar wind bubbles within WMAP
sources. For G49, the dynamical timescale estimated from the expansion
of these bubbles is $\sim$ 2 Myr. We investigated using Starburst99
the evolution of a $7.5 \times 10^3$ \Msun\ cluster and 
the change of $Q_{He}/Q_{H}$ with its age. Fig.~\ref{fig2} shows the result. 
The ratio $Q_{He}/Q_{H}$ decreases by a factor of 3.4 by $\sim$ 2.5 Myr
and then increases due to the evolution of massive stars to Wolf-Rayet (WR) stars.
The surface temperatures of WR stars are $>$ 50,000 K and hence they
contribute significantly to He ionization. Note that by about 2.8 Myr or so the
total Lyc flux due to the cluster drops down. As can be seen from Fig~\ref{fig2}b, 
the reduction in $Q_{He}/Q_{H}$ at age $\sim$ 2 Myr is not sufficient to explain 
the observed upper limit of He line.

The observed limit to the He line can be used to obtain an upper limit
on the ratio $Q_{He}/Q_{H}$ using the He ionization model of
\nocite{m71}Mathis (1971). The upper limit thus obtained for $Q_{He}/Q_{H}$ 
is about 0.028, which is marked
on Fig~\ref{fig2}b. We modified Starburst99's massive star atmospheric 
model emission routines to introduce an attenuation for photons with energy $\ge$ 
24.6 eV. These photons need to be attenuated by at least a factor of 10
relative to the current model results if the ionizing sources are 
star clusters with age \lsim\ 2.8 Myr (see Fig.~\ref{fig2}b). 

Environments of stellar clusters and their evolution have been recently 
studied both theoretically (\nocite{pz11}Pelupessy \& Zwart 2011)
and observationally (\nocite{cetal06}Churchwell et al 2006).
These studies show that the cluster wind is a dominant factor affecting
the environment at all stages of cluster evolution. These winds can
produce dust bubbles. We follow the treatment of \nocite{metal74}Mezger \etal\ (1974;
see also \nocite{ps78}Panagia \& Smith 1978), developed for compact \HII\ regions, 
to investigate whether selective absorption of He-ionizing photons by dust 
can explain the non-detection of He RRL toward the DIR. The size of G49 is
$\sim$ 70 pc, which along with electron density ($\sim$ 5\cmthree;
\nocite{ra01}Roshi \& Anantharamaiah 2001) gives a dust optical depth 
at $\lambda$ = 912 \AA\ in the range 0.25 to 0.8 depending on the 
absorption cross section at this wavelength. The dust optical depth 
near 0.25 cannot significantly change $n_{He^+}/n_{H^+}$ 
in \HII\ regions (\nocite{s77}Sarazin 1977).
Using the observed limit on $n_{He^+}/n_{H^+}$, $Q_{He}/Q_{H} \sim 0.24$ 
and the above estimated dust optical depth, we find that the ratio of the dust
absorption cross sections for He- and H-ionizing photons is \gsim 6. 
This high cross section ratio is inconsistent with our current
understanding of UV absorption properties of the dust (\nocite{d03}Draine 2003). 

\section{Summary}
\label{sum}

Radio recombination line observations of the luminous WMAP source G49
were made at 1.4 GHz with the GBT. Hydrogen and carbon lines were
detected but no helium line was observed. The upper limit to the ratio
of ionized helium to hydrogen obtained is 0.024. It has been suggested that the
WMAP sources are ionized by star clusters which may not be associated with
the radio \HII\ regions within these sources. Using Starburst99 we found that the mass of the
cluster should be $\sim 7.5 \times 10^3$ \Msun\ to satisfy the total ionization 
requirement of G49. The ratio $Q_{He}/Q_H$ obtained for such a cluster
is $\sim 0.24$, implying that the hydrogen and helium Str\"{o}mgren spheres should
overlap and hence the helium line should be detectable. The non-detection 
of helium line rules out the possibility that the DIR is an \HII\ region 
produced by such stellar clusters. We examined whether
the DIR is ionized by `leaky' \HII\ regions embedded in G49, as
suggested by \nocite{l76} Lockman (1976) and \nocite{a86} Anantharamaiah (1986). 
Models of `leaky' \HII\ regions 
show that the helium ionizing photons are significantly suppressed for 
photon leakage \lsim\ 15\% and for ionizing source with effective temperature \lsim\ 45000 K; 
both these conditions are not met for G49. We determined that photons
with energy $\ge$ 24.6 eV need to be attenuated by at least a factor of
10 compared to the current model results to be consistent with our observations. 
If selective absorption of He-ionizing photons by dust is causing this additional
attenuation, then the ratio of dust absorption cross sections for He- and H-ionizing 
photons should be \gsim 6. 

\section{Acknowledgments}

DAR thanks J. Lockman and R. Maddalena for helpful discussions during the planning
stage of the GBT observations. DAR also thanks Ed Churchwell and Claus Leitherer 
for informative discussion during the interpretation of the results. 
The data were taken as part of a project for the Single dish school, held at NRAO, 
Green Bank during July 2011.

{\it Facility:} \facility{GBT}

{}

\clearpage

\clearpage

\begin{deluxetable}{lrrr}
\small
\tablecaption{Parameters of the observed Recombination Lines toward G49.75$-$0.45 \label{tab1}}
\tablewidth{0pt}
\tablehead{
\colhead{} & \colhead{$T_{\rm L}$} & \colhead{$V_{\rm LSR}$} & \colhead{$\Delta{V}$} \\
\colhead{Transition} & \colhead{(K)} & \colhead{(\kms)} & \colhead{(\kms)}
}
\startdata
H$^1$  &   0.0324(0.0003) & 60.46(0.15)    & 28.95(0.35) \\
He &   (0.00078)$^{2}$      &  \nodata       &  \nodata    \\
C$^1$  &   0.0048(0.0009) & 69.83(0.37)    &  3.87(0.91) \\
C$^1$  &   0.0050(0.0012) & 64.35(0.27)    &  2.18(0.64) \\
X$^{1,3}$  &   0.0081(0.0016) & $-$48.21(0.12) &  1.26(0.29) \\
\enddata
\tablenotetext{1}{The line parameters are obtained from the spectrum smoothed
to a velocity resolution of 1.1 \kms.} 
\tablenotetext{2}{See Section~\ref{result} for the determination of upper limit to He 
line emission. } 
\tablenotetext{3}{This transition is not identified. The listed LSR velocity of X 
line is with respect to hydrogen.} 
\end{deluxetable}

\end{document}